# Status of the NEMO Project

Piera Sapienza for the NEMO collaboration

*Abstract*—The NEMO project aims at the search, development and validation of key technologies for the construction, deployment and mantainance of an underwater Cherenkov $km^3$ neutrino telescope in the Mediterranean Sea. Moreover, the NEMO Collaboration carried out a long term exploration of a 3500 m deep sea site close to the Sicilian coast; the study has shown that it is optimal for the installation of the detector. A Phase-1 project, which is under way, will validate the proposed technologies for the km3 detector on a Test Site at 2000 m depth. The realization of a new infrastructure on the candidate site (Phase-2 project) will provide the possibility to test detector components at 3500 m depth and will allow also a continuous monitoring of the site.

## I. Introduction

THE expectations on neutrino fluxes based on the measured cosmic ray fluxes [1], as well as one estimate of probable neutrino fluxes from galactic γ-TeV sources [2] indicate that the opening of the high energy muon neutrino astronomy era can only be performed with detectors of $km^3$ scale.

A first generation of small scale detectors has been realized (AMANDA [3] at the South Pole and NT-200 [4] in the Baikal lake) and have set limits on neutrino fluxes, while others are at different stage of realization (ANTARES [5] and NESTOR [6]). Following the success of AMANDA, which is presently the largest operating detector, the realization of the IceCube $km^3$ detector [7] has started at the South Pole. On the other hand, the necessity of full sky coverage strongly support the complementary construction of a $km^3$-scale detector in the Mediterranean Sea.

The activity of the NEMO collaboration has been mainly focused on the search and characterization of an optimal site for the detector installation and on the development of key technologies for the $km^3$ underwater telescope.

Piera Sapienza is with the Istituto Nazionale di Fisica Nucleare - Laboratori Nazionali del Sud, Via S. Sofia 62, I-95123 Catania, Italy (corresponding author to provide phone: #39-095-542288; e-mail: sapienza@lns.ifn.it).

The feasibility study of the $km^3$ detector includes the analysis of all the construction and installation issues and the optimization of the detector geometry by means of numerical simulations. The validation of the proposed technologies via an advanced R&D activity, the prototyping of the proposed technical solutions and their validation in a deep sea environment will be carried out with the two pilot projects NEMO Phase-1 and Phase-2.

## II. Site exploration

The Mediterranean Sea offers optimal conditions, on a worldwide scale, to locate the $km^3$ telescope. The seabed coast can reach depths beyond 3000 m, at distances less than a hundred kilometers from the shore. These characteristics are very important, since depth helps to filter out the bulk of the low energy component of down going atmospheric muons and the relatively close distance to the coast allows the data transfer from the detector to shore by means of standard commercial electro-optical cables.

After 30 sea campaigns since July 1998, a site located in the Ionian Sea (36° 19' N, 16° 05' E), close to Capo Passero in the South-East part of Sicily, was identified as the best candidate. The site is a wide abyssal plateau with an average depth of about 3500 m, located at less than 80 km from the shore and about 50 km from the shelf break.

Water transparency was measured *in situ* using a set-up based on a transmissometer that allowed to measure light absorption and attenuation at nine different wavelengths ranging from 412 to 715 nm [8]. The values of the absorption length measured at depths of interest for the $km^3$ detector installation (more than 2500 m) are close to the one of optically pure sea water (about 70 m at $\lambda$ = 440 nm). Seasonal variations are negligible and compatible with the instrument experimental error [9].

Another characteristic of the deep sea water that can affect the detector performance is the optical background. This background comes from two natural causes: the decay of $^{40}K$, which is present in seawater, and the so called *bioluminescence* that is the light produced by biological entities. Of these two effects the first one shows up as a constant rate background noise on the optical modules, while the second one, when present, may induce large fluctuations (both in the baseline and as high rate spikes) in the noise rate.

In Capo Passero an average rate of about 30 kHz of optical noise, compatible with what expected from pure $^{40}$K background, with rare high rate spikes due to bioluminescence was measured at a depth of 3000 m in several sea campaigns. This result is in agreement with the vertical distribution of bioluminescent bacteria measured in Capo Passero, that shows a very low concentration of these bacteria at depths greater than 2500 m.

Current meter chains were moored in the region of Capo Passero since July 1998, to measure the deep sea current intensity and direction. Over more than 7 years of data taking the currents at a depth of about 3000 m appear to be low and regular (2-3 cm/s average; 12 cm/s maximum).

The downward flux of sediments has also been analyzed. The annual average value of material sedimenting at large depth in Capo Passero is about 60 mg m$^{-2}$ day$^{-1}$, which is a rather small value, this was expected for an oligotrophic environment such as the Ionian Plateau.

Altogether the features of the Capo Passero site appear to be optimal for the installation of an underwater km3 neutrino telescope. The Capo Passero site was proposed as candidate site to host the Mediterranean km3 neutrino telescope to APECC in January 2003.

### III. DETECTOR LAYOUT AND PERFORMANCE

The design of an underwater km$^3$ neutrino telescope represents a challenging task that has to match many requirements concerning the detector performances, the technical feasibility and the project budget. In general, a km$^3$ detector is an array of vertical structures each one hosting several tens of optical modules (OM) arranged in a non-homogenous distribution. Moreover, it is important to limit the total number of structures in order to reduce the number of underwater connections.

The detector performances were evaluated by means of numerical simulations, carried out using the software [10] developed by the ANTARES collaboration and adapted to km$^3$ scale detectors [11].

The proposed NEMO architecture is a $9 \times 9$ square lattice of towers. Details on the structure of the tower will be given in section 4.3. Here we just mention that in the simulations a configuration with 18 floors, each one hosting four OMs with 10" PMT was used for a total of 5832 PMTs.

The proposed architecture is "modular" and the layout can be reconfigured to match different detector specifications.

Site dependent parameters such as depth, optical background, absorption and scattering length, have been set accordingly with the values measured in Capo Passero at a depth of about 3400 m.

The sensitivity of the detector for a generic astrophysical point-like source is reported in fig. 1 as a function of the integrated data taking time. The source position was chosen at a declination δ = -60°; an E$^{-2}$ neutrino energy spectrum was considered. For a comparison the IceCube sensitivity [12] obtained for a 1° search bin is also reported in the figure.

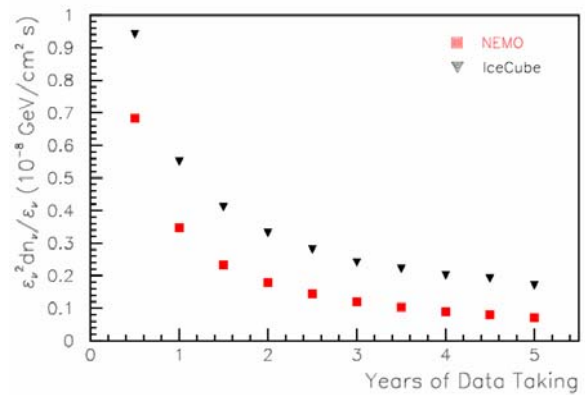

Figure 1. Sensitivity of the NEMO km3 detector to an astrophysical neutrino point-like source with declination δ = -60° and a E$^{-2}$ energy spectrum compared to the IceCube sensitivity [11].

A simulation study of the Moon shadow effect, due to the absorption of primary cosmic rays by the Moon, has also been undertaken. Indeed, this effect should provide a direct measurement of both the detector angular resolution and pointing accuracy.

Preliminary results, presented in [13], show an angular resolution s = 0.19° ± 0.02°.

The energy range of interest for high energy neutrino telescopes is very broad, spanning from few hundreds of GeV to very high energies such as the GZK energies (~ $10^{20}$ eV). Therefore, the detector layout should be optimized with respect to the physics one wants to focus on. The possibility to reconfigure the detector layout to tailor to different detection needs is a specific feature of underwater detectors.

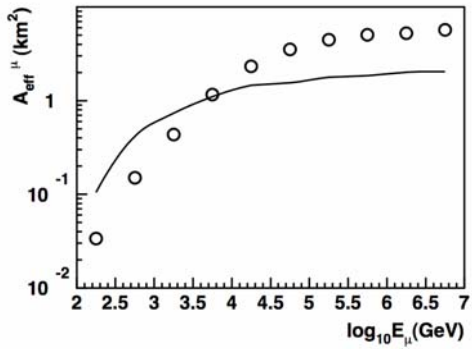

Figure 2. Effective areas for two different configurations of the NEMO $km^3$ detector: 140 m tower interspacing (solid line) and 300 m tower interpacing (circles).

The effective area for the detection of very energetic muon neutrino ($E_\nu \geq 100$ TeV) can be enhanced by increasing the distance between structures. In fig. 2 the effective areas as a function of the muon energy are reported for two 9 × 9 detector configurations, with the same number of structures and OMs, for two different inter-tower distances of 140 m and 300 m. A significant gain is observed for the sparser array at the expenses of a higher energy detection threshold.

## IV. THE NEMO PHASE 1 PROJECT

As an intermediate step towards the underwater $km^3$ detector and to ensure an adequate process of validation, the realization of a technological demonstrator was considered. This project, called NEMO Phase-1, includes prototypes of the critical elements of the proposed $km^3$ detector: the junction box and the tower. This will allow to validate the mechanical characteristics of both as well as the data transmission and power distribution system of the whole apparatus.

The project is under realization at the Underwater Test Site of the Laboratori Nazionali del Sud in Catania, where a 28 km electro optical cable, reaching the depth of 2000 m, allows the connection of deep sea intrumentation to a shore station.

The Test Site of the Laboratori Nazionali del Sud consists of an underwater installation at 2000 m depth, connected by means of an electro-optical submarine cable the to shore, and a shore station.

The cable system is composed of a 23 km main electro-optical cable, split at the end in two branches, each one 5 km long. One branch is used for the NEMO Phase-1 experiment, while the other provides connection for the SN-1 underwater seismic monitoring station, realized by the Istituto Nazionale di Geofisica e Vulcanologia (INGV) [14]. The cable carries 6 electrical wires and 10 monomode optical fibres.

A shore station, located inside the port of Catania, hosts the power system, the instrumentation control system, the land station of the data transmission system and the data acquisition.

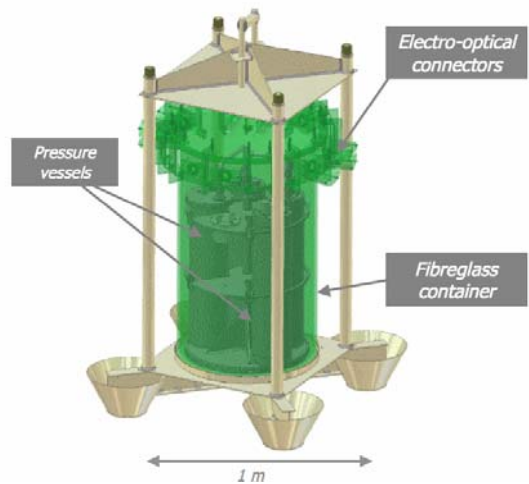

Figure 3. The NEMO Phase-1 Junction Box.

The Junction Box (JB) is a key element of the

detector. It must provide connection between the main electro-optical cable and the detector structures and has to be designed to host and protect from the effects of corrosion and pressure, the opto-electronic boards dedicated to the distribution and the control of the power supply and digitized signals.

An alternative design to the standard Titanium pressure vessels used for junction boxes operating in seawater for a long time has been developed. The approach followed is to decouple the two problems of pressure and corrosion resistance. The JB is realized with four cylindrical steel vessels hosted in a large fibreglass container to avoid direct contact between steel and sea water (Fig. 3). The fibreglass container is filled with silicone oil and pressure compensated. All the electronics able to withstand high pressure is installed in oil bath, while the rest is located inside one pressure resistant container [15].

The tower that will host the optical modules and the instrumentation is a three dimensional flexible structure composed by a sequence of floors (that host the instrumentation) interlinked by cables and anchored on the seabed [16]. The structure is kept vertical by appropriate buoyancy on the top.

While the design of a complete tower for the km$^3$ foresees 16 floors, the prototype under realization for the Phase-1 project is a "mini-tower" of 4 floors, each made with a 15 m long structure hosting two optical modules (one down-looking and one horizontally-looking) at each end (4 OM per storey). The floors are vertically spaced by 40 m. Each floor is connected to the following one by means of four ropes that are fastened in a way that forces each floor horizontal structure to take an orientation perpendicular with respect to the adjacent (top and bottom) ones. An additional spacing of 150 m is added at the base of the tower, between the tower base and the lowermost floor to allow for a sufficient water volume below the detector.

A scheme of the prototype four floor tower is shown in fig. 4. In addition to the 16 Optical Modules the instrumentation installed includes several sensors for calibration and environmental monitoring. In particular two hydrophones are mounted on the tower base and at the extremities of each floor. These, together with an acoustic beacon placed on the tower base and other beacons installed on the sea bed, are used for precise determination of the tower position by means of time delay measurements of acoustic signals. The other environmental probes are: a Conductivity-Temperature-Depth (CTD) probe used for monitoring of the water temperature and salinity, a light attenuation probe (C*) and an Acoustic Doppler Current Profile (ADCP) that will provide continuous monitoring of the deep sea currents along the whole tower height.

The design of the cabling system [17] of the tower was made taking into account the fault tolerance and the ease of management. The goal has been the minimization of the connection interfaces.

At the base of the tower there is a Tower Base Module (TBM). The TBM is connected by means of a hybrid penetrator to the backbone cabling and by means of a hybrid connector to a jumper cable terminated with a wet mateable bulkhead that allows the interconnection of the tower to the Junction box by means of a ROV.

A lightweight electro-optical backbone cabling system distributes the power and allows the data transmission signals to and from the tower floors. The splitting of the cable is performed by means of breakouts positioned at each floor level. The breakouts are realized with pressure vessels each one containing two passive optical devices that perform the Add/Drop functions for the optical data transmission signals of the outgoing and incoming data.

Inside each floor structure two containers are installed: a Floor Power Module (FPM) and a Floor Control Module (FCM). This last one is the core of the system since it hosts all the floor electronics for data transmission. This module is realized with a solution analogous to that adopted for the Junction Box: a metallic pressure resistant vessel placed inside an external plastic container filled with silicone oil and pressure compensated. The FPM is a silicone oil filled plastic container; the power supply subsystem that is hosted inside has been tested to operate under pressure [16]. The FCM is interfaced to the floor instrumentation by means of four electro-optical (for the OMs) and three electrical (for the additional

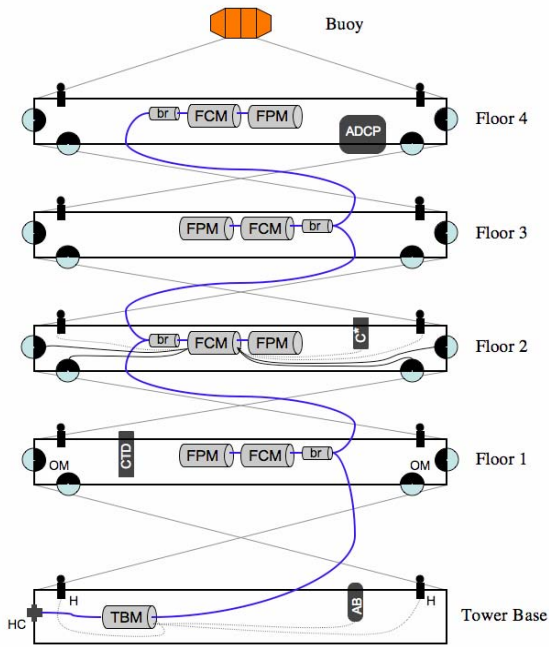

Figure 4. Scheme of the four floor prototype tower of the NEMO Phase-1 project. The instrumentation mounted on it includes: 16 Optical Modules (OM); 10 Hydrophones (H); 1 Acoustic Beacon (AB) on the Tower Base; 1 Current-Temperature-Depth probe (CTD) on the first floor; 1 probe for light attenuation measurements (C*) on the second floor; 1 Acoustic Doppler Current Profiler (ADCP) on the fourth floor. The scheme of the backbone cabling including the Tower Base Module (TBM), the floor breakouts (br), the Floor Control Modules (FCM) and the Floor Power Modules (FPM) is shown. Connection to the Junction Box is provided through a wet mateable hybrid connector (HC) placed on the tower base. For clarity the layout of the floor internal cabling is drawn only for floor 2, with electro-optic connections as continuous lines and electric connection as dotted lines.

instrumentation) penetrators.

In this cabling system connectors are positioned at the subsystems interfaces, to allow for testing of each single subsystem and for ease of assembly, and at users interfaces only. This allows to reduce their number thus reducing the cost of the system and increasing its reliability. Moreover, the use of penetrators instead of connectors minimizes the optical losses allowing for a higher optical power budget for the data transmission system.

The optical module is essentially composed by a photo-multiplier (PMT) enclosed in a 17" pressure resistant sphere of thick glass. The used PMT is a 10" Hamamatsu R7081Sel with 10 stages.

In spite of its large photocathode area, the Hamamatsu PMT R7081Sel has a good time resolution of about 3 ns FWHM for single photoelectron pulses with a charge resolution of 35%.

Mechanical and optical contact between the PMT and the internal glass surface is ensured by an optical silicone gel. A μ-metal cage shields the PMT from the Earth's magnetic field.

The base card circuit for the high voltage distribution (Iseg PHQ 7081SEL) requires only a low voltage supply (+5 V) and generates all necessary voltages for cathode, grid and dynodes with a power consumption of less than 150 mW.

A front-end electronics board, built with discrete components, was designed, realized and tested [18]. This board is placed inside the OM. Sampling at 200 MHz is accomplished by two 100 MHz staggered Flash ADCs, whose outputs are captured by an FPGA which classifies (according to a remotely programmable threshold) the signal as valid or not; stores it with an event time stamp in an internal 12 kbit FIFO; packs OM data and local slow control information; and codes everything into a bit stream frame ready to be transmitted on a differential pair at 20 Mbit/s rate. The main features of this solution are the moderate power consumption, the high resolution, the large input dynamic range obtained by a quasi-logarithmic analog compression circuit and the fine time resolution. Through an incoming slow control channel, managed by a DSP, all the acquisition parameters can be changed, and there is the possibility to remotely re-program the FPGA downloading new codes. Moreover, the board contains electronics, analog and digital, to control the Optical Module power supply to measure temperature, humidity and electrical parameters and to auto calibrate the non linear response of the logarithmic compressor.

The design of the data transport system for NEMO Phase-1 was based on technical choices that allow scalability to a much bigger apparatus [19]. For synchronization purposes, a common timing must be known in the whole apparatus at the level of detection device to allow correlation in time of events. For this reason the Synchronous Digital Hierarchy (SDH) protocol, which embeds data, synchronism and clock timing in the same serial bit stream, and allows an easy distribution of the clock signal to the whole apparatus, has been chosen. For communication the technology adopted relies on Dense Wavelength Division Multiplex (DWDM) techniques, using totally passive components with the only exception of the line termination devices, i.e. electro-optical transceivers. The great advantages in terms of power consumption, reliability, and simplicity recommend this technique as a perfect candidate for final $km^3$ detector.

The FCM on each floor collects data from the floor OMs and the auxiliary instrumentation, creates an STM-1/SDH data stream at about 155 Mbps, and sends data toward the shore laboratory. From the opposite direction, the FCM receives slow control data, commands and auxiliary information, and the clock and synchronization signals needed for apparatus timing. Bidirectional data transport is realized by means of the backbone optical fiber cabling system described in sec. 4.4. In order to provide redundancy, data streams are doubled and re-directed onto two fibers using a "power splitter". The fiber, of the two, that carries the meaningful information is chosen at on-shore station.

The underwater structure has a mirrored on-shore counterpart, where all optical signals are reconverted into electrical signals. In the on-shore laboratory the Primary Reference Clock, which is used to give the same timing to all the towers of the apparatus, is also located. Assuming that the two fibers per tower maintain their integrity, the designed system provides other experiments with a further bidirectional channel.

For the Phase-1 project a three phase AC system has been chosen since it presents some advantages in terms of voltage drops and reliability. This system is used for the energy distribution up to the level of the local electronics module in each storey where a conversion to DC is made [16].

A control system was realized to acquire all the relevant data such as currents, voltages and environmental parameters (temperature, humidity, etc…) inside the containers.

The system has been designed to have a large part of its components working under pressure inside an oil bath. For this aim extensive tests on electric and electronics components have been performed [16].

The timing calibration requires an embedded system in order to track the possible drifts of the time offsets during the operations of the apparatus underwater. The task of this embedded system is essentially to measure the offsets with which the local time counters inside the optical modules are reset on reception of the reset commands sent from shore, i.e. the time delays for such commands to reach the individual optical modules. All time measurements are in fact referred to the readout of such counters. The operation will be performed with a completely redundant system [20]: 1) a two-step procedure for measuring the offsets in the time measurements of the optical sensors; 2) an all-optical procedure for measuring the differences in the time offsets of the different optical modules.

In the first system the needed measurements are performed in two separate steps:
- with an 'echo' timing calibration;
- with an 'optical' timing calibration.

The former will allow to measure the time delay for the signal propagation from shore to the FCM of each floor; the latter, which is based on a network of optical fibres which distributes calibration signals from fast pulsers to the optical modules, will allow to determine the time offsets between the FCM and each optical module connected to it. The second system is an extension of the optical timing calibration system, which allows to simultaneously calibrate the optical modules of different floors of the tower.

An essential requirement for the muon tracking is the knowledge of each sensor position. While the position and orientation of the tower base is fixed and known from its installation, the rest of the structure can bend under the influence of sea currents [16]. A precise determination of the position of each tower floor is achieved by means of triangulations performed by measuring time delays of acoustic signals between acoustic beacons placed on the sea floor and a couple of hydrophones installed on each tower floor. In addition to this the inclination and orientation of each tower floor is measured by a tiltmeter and a compass placed inside the FCM.

All the Slow Control data (including data from all environmental sensors and the acoustic positioning system) are managed from shore by means of a dedicated Slow Control Management System [21].

### A. Project status

The submarine cable was deployed in 2001. In January 2005 the project has achieved a major milestone with the installation of the two cable termination frames (that host the wet mateable underwater bulkheads) and the deployment and connection of a small acoustic detection station (for the detection of acoustic background in the deep sea environment) [22] on one branch and the SN-1 seismic and environmental observatory on the other branch [14]. Both stations have been operational ever since, continuously transmitting data to shore.

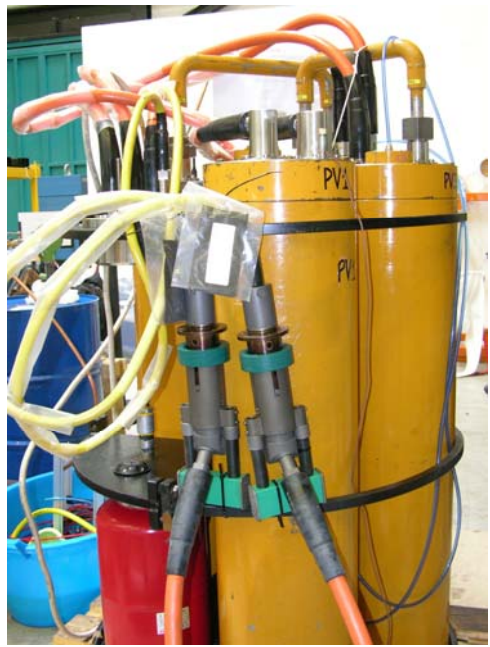

Figure 5. The fully assembled internal vessels of the Junction Box.

All the Phase-1 detector subsystems have undergone laboratory tests for their qualification. The data transmission system has been extensively and successfully tested with particular emphasis on the photonic circuit that have to assure an adequate optical power budget for the data transport. Also the power feeding and control system has been tested in its complete configuration.

The integration of the junction box (Fig. 5) is completed while the mini-tower (Fig. 6) integration is almost completed. Their deployment and connection is scheduled forthe autumn of 2006.

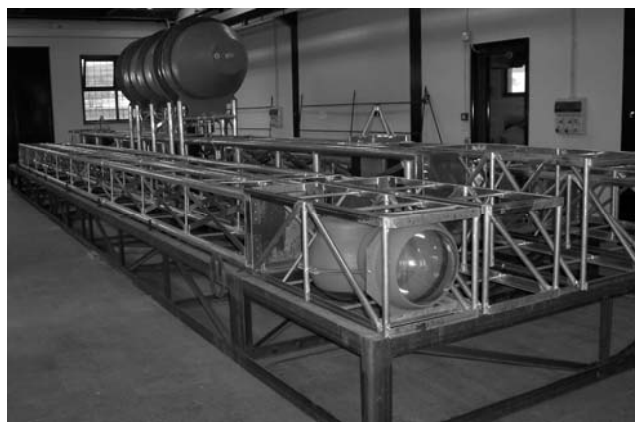

Figure 6. Integration of the tower structure.

## V. THE NEMO PHASE 2 PROJECT

Although the Phase-1 project will provide a fundamental test of the technologies proposed for the realization and installation of the detector, these must be finally validated at the depths needed for the km$^3$ detector.

For these motivations the realization of an infrastructure at the Capo Passero site has been undertaken. It will consist of a 100 km cable, linking the 3500 m deep sea site to the shore, a shore station, located inside the harbour area of Portopalo di Capo Passero, and the underwater infrastructures needed to connect prototypes of the km$^3$ detector.

Due to the longer cable needed, a different solution, with respect to the Phase-1 project, for the electro-optical cable was chosen. In this case the backbone cable will be a DC cable, manufactured by Alcatel, that will carry a single electrical conductor, that can be operated at 10 kV DC allowing a power transport of more than 50 kW, and 20 single mode optical fibres for data transmission [24].

The completion of this project is foreseen by the end of 2007. In that time it will be possible to connect one or more prototypes of detector structures, allowing a full test at 3500 m of the deployment and connection procedures. This project will also allow a continuous long term on-line monitoring of the site properties (light transparency, optical background, water currents, …) whose knowledge is essential for the installation of the full km$^3$ detector.

VI. CONCLUSION

The realization of an underwater km$^3$ telescope for high energy astrophysical neutrinos requires the development and validation of technologies appropriate for a hostile environment such as the deep ocean. The NEMO collaboration works in this direction through an intense R&D activity.

An extensive study on a 3500 m deep site close to the coast of Sicily has demonstrated that it has optimal characteristics for the telescope installation and the site was candidate to APECC in January 2003 for the installation of the underwater Cherenkov km$^3$ neutrino telescope in the Mediterranean Sea.

A demonstrator of the technological solutions proposed for the km$^3$ detector is under final assembly at the shore station of the LNS Test Site in the Catania harbor. The deployment and connection of the junction box and the minitower at the 2000 m depth offshore infrastructure of the underwater Test Site is foreseen for November 2006.

Moreover, a Phase-2 project, which aims at the realization of a new infrastructure on the deep sea site of Capo Passero (3500 m), started.

A further R&D program will be developed within the KM3NeT Design Study [25] in which all the European institutes currently involved in the Mediterranean neutrino astronomy projects are participating. The project, partly supported by the European Union, started in February 2006 and aims at producing a Technical Design Report for the realization of an underwater Cherenkov km$^3$-scale neutrino telescope in the Mediterranean Sea in three years.